# Adjusting PageRank parameters and Comparing results

Subhajit Sahu[1], Kishore Kothapalli[1], Dip Sankar Banerjee[2]
[1]International Institute of Information Technology, Hyderabad
[2]Indian Institute of Technology, Jodhpur

**Abstract** — The effect of adjusting *damping factor α* and *tolerance τ* on *iterations* needed for PageRank computation is studied here. Relative performance of PageRank computation with *L1*, *L2*, and *L∞ norms*  used as convergence check, are also compared with *six* possible *mean ratios*. It is observed that *increasing* the *damping factor α* linearly *increases* the *iterations* needed *almost exponentially*. On the other hand, *decreasing* the *tolerance τ* exponentially *decreases* the *iterations* needed *almost exponentially*. On average, PageRank with *L∞ norm* as convergence check is the *fastest*, quickly followed by *L2 norm*, and then *L1 norm*. For large graphs, above certain *tolerance τ* values, convergence can occur in a *single iteration*. On the contrary, below certain *tolerance τ* values, *sensitivity issues* can begin to appear, causing computation to halt at maximum iteration limit without convergence. The six mean ratios for relative performance comparison are based on arithmetic, geometric, and harmonic mean, as well as the order of ratio calculation. Among them GM-RATIO, geometric mean followed by ratio calculation, is found to be most stable, followed by AM-RATIO.

**Index terms** — PageRank algorithm, Parameter adjustment, Convergence function, Sensitivity issues, Relative performance comparison.

---

## 1. Introduction

Web graphs unaltered are reducible, and thus the **rate of convergence** of the *power-iteration method* is the rate at which $α^k → 0$, where $α$ is the *damping factor*, and $k$ is the *iteration count*. An estimate of the number of iterations needed to converge to a *tolerance τ* is $log_{10} τ / log_{10} α$ [1]. For $τ = 10^{-6}$ and $α = 0.85$, it can take roughly *85 iterations* to converge. For $α = 0.95$, and $α = 0.75$, with the same *tolerance* $τ = 10^{-6}$, it takes roughly *269* and *48 iterations* respectively. For $τ = 10^{-9}$, and $τ = 10^{-3}$, with the same *damping factor* $α =$

*0.85*, it takes roughly *128* and *43 iterations* respectively. Thus, adjusting the *damping factor* or the *tolerance parameters* of the PageRank algorithm can have a significant effect on the *convergence rate*.

However, once results for various test cases are obtained, there exist multiple methods to obtain a **composite relative performance ratio**. Consider, for example, three approaches *a*, *b*, and *c*, with 3 test runs for each of the three approaches, labeled $a_1, a_2, a_3, b_1, b_2, b_3, c_1, c_2, c_3$. One method to get a composite ratio between the three approaches would be to find the relative *performance ratio* of *each approach* with respect to a *baseline approach* (one of them), and then calculate the *arithmetic-mean (AM)* for each approach. For example, the relative performance of each approach with respect to *c* would be $a_1/c_1$, $b_1/c_1$, $c_1/c_1$, $a_2/c_2$, $b_2/c_2$, and so on. The RATIO-AM with respect to *c* is now the *arithmetic mean* of these ratios, i.e., $(a_1/c_1 + a_2/c_2 + a_3/c_3)/3$ for *a*, $(b_1/c_1 + b_2/c_2 + b_3/c_3)/3$, and *1* for *c*. Similarly, RATIO-GM, and RATIO-HM can be obtained by instead calculating *geometric mean (GM)*, or *harmonic mean (HM)* respectively. Unfortunately, based upon the choice of the baseline approach, the composite ratios can differ (except RATIO-GM, as discussed later). The alternative approach is to calculate means for each approach first, and then find the relative performance ratio. Like before, arithmetic, geometric, or harmonic mean can be used, called AM-RATIO, GM-RATIO, and HM-RATIO respectively. For example, the arithmetic mean of each approach would be $(a_1+a_2+a_3)/3$ for *a*, $(b_1+b_2+b_3)/3$ for *b*, and so on. The AM-RATIO with respect to *c* is now the ratio of means of each approach with respect to *c*'s, i.e., $(a_1+a_2+a_3)/(c_1+c_2+c_3)$ for *a*, $(b_1+b_2+b_3)/(c_1+c_2+c_3)$ for *b*, and *1* for *c*. Since the ratio is calculated last, the choice of the baseline approach has no effect on the result. Note that, GM-RATIO and RATIO-GM give identical composite ratios due to the mathematical property of geometric mean, and thus the order of ratio calculation does not affect the result. Weighted geometric mean has been selected for SPECviewperf® composite numbers [2].

## 2. Adjusting Damping factor

Adjustment of the *damping factor* $α$ is a delicate balancing act. For smaller values of $α$, the convergence is fast, but the *link structure of the graph* used to determine ranks is less true. Slightly different values for $α$ can produce *very different* rank vectors. Moreover, as α → 1, convergence *slows down drastically*, and *sensitivity issues* begin to surface [1].

For this experiment, the **damping factor α** (which is usually *0.85*) is **varied** from *0.50* to *1.00* in steps of *0.05*. This is in order to compare the performance variation with each *damping factor*. The calculated error is the *L1 norm* with respect to default PageRank (*α = 0.85*). The PageRank algorithm used here is the *standard power-iteration (pull) based PageRank*. The rank of a vertex in an iteration is calculated as **$c_0$ + α∑$r_n$/$d_n$**, where $c_0$ is the *common teleport contribution*, *α* is the *damping factor*, $r_n$ is the *previous rank of vertex* with an incoming edge, $d_n$ is the *out-degree* of the incoming-edge vertex, and *N* is the *total number of vertices* in the graph. The *common teleport contribution* **$c_0$**, calculated as **(1-α)/N + α∑$r_n$/N**, includes the *contribution due to a teleport from any vertex* in the graph due to the damping factor *(1-α)/N*, and *teleport from dangling vertices* (with *no outgoing edges*) in the graph *α∑$r_n$/N*. This is because a random surfer jumps to a random page upon visiting a page with *no links*, in order to avoid the *rank-sink* effect.

All *seventeen* graphs used in this experiment are stored in the *MatrixMarket (.mtx)* file format, and obtained from the *SuiteSparse Matrix Collection*. These include: *web-Stanford, web-BerkStan, web-Google, web-NotreDame, soc-Slashdot0811, soc-Slashdot0902, soc-Epinions1, coAuthorsDBLP, coAuthorsCiteseer, soc-LiveJournal1, coPapersCiteseer, coPapersDBLP, indochina-2004, italy_osm, great-britain_osm, germany_osm, asia_osm*. The experiment is implemented in *C++*, and compiled using *GCC 9* with *optimization level 3 (-O3)*. The system used is a *Dell PowerEdge R740 Rack server* with two *Intel Xeon Silver 4116 CPUs @ 2.10GHz, 128GB DIMM DDR4 Synchronous Registered (Buffered) 2666 MHz (8x16GB) DRAM*, and running *CentOS Linux release 7.9.2009 (Core)*. The *iterations* taken with each test case is measured. *500* is the *maximum iterations* allowed. Statistics of each test case is printed to *standard output (stdout)*, and redirected to a *log file*, which is then processed with a *script* to generate a *CSV file*, with each *row* representing the details of a *single test case*. This *CSV file* is imported into *Google Sheets*, and necessary tables are set up with the help of the *FILTER* function to create the *charts*.

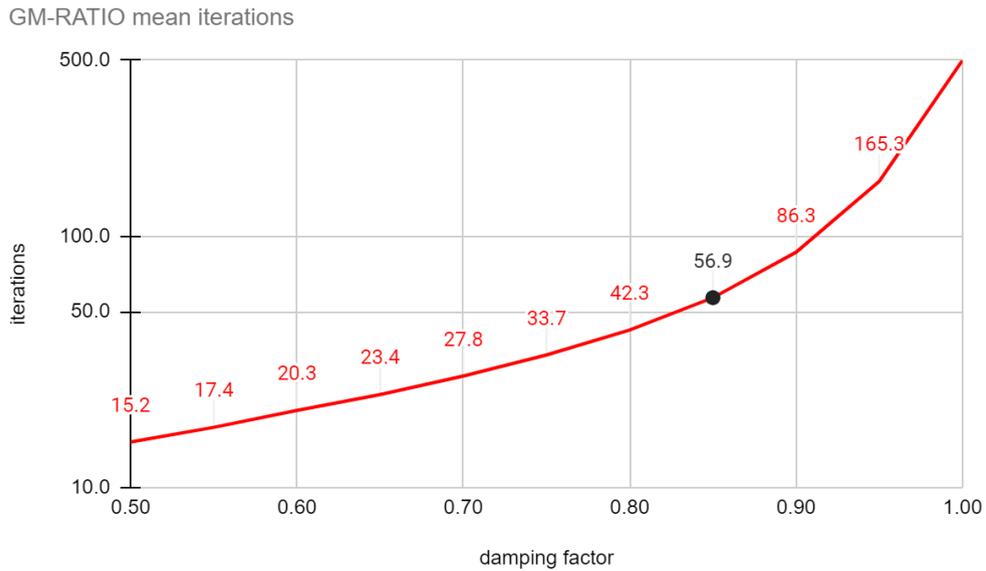

Figure 2.1: Geometric mean iterations for PageRank computation with *damping factor α* adjusted from *0.50 - 1.00* in steps of *0.05*. Chart for AM iterations is quite similar.

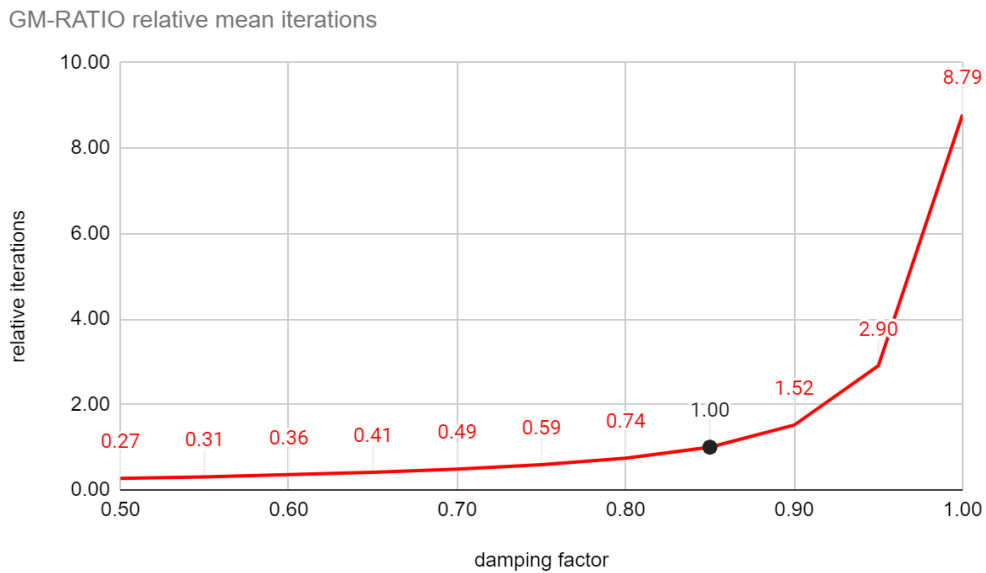

Figure 2.2: Relative geometric mean (GM-RATIO) iterations for PageRank computation with *damping factor α* adjusted from *0.50 - 1.00* in steps of *0.05*. Chart for AM-RATIO is quite similar.

Results, as shown in figures 2.1 and 2.2, indicate that **increasing the damping factor α beyond *0.85* significantly increases convergence time**, and lowering it below *0.85* decreases convergence time. As the *damping factor α* increases *linearly*, the iterations needed for PageRank computation increases *almost exponentially*. On average, using a *damping factor α = 0.95 increases iterations* needed by *190% (~2.9x)*, and using a *damping factor α = 0.75 decreases* it by *41% (~0.6x)*, compared to *damping factor α = 0.85*. Note that a higher *damping factor* implies that a random surfer follows links with *higher probability* (and jumps to a random page with lower probability).

## 3. Adjusting Error function

It is observed that a number of *error functions* are in use for checking convergence of PageRank computation. Although *L1 norm* is commonly used for convergence check, it appears *nvGraph* uses *L2 norm* instead [3]. Another person in stackoverflow seems to suggest the use of *per-vertex tolerance comparison*, which is essentially the *L∞ norm* [4]. The **L1 norm** $||E||_1$ between two *(rank) vectors r* and *s* is calculated as $\sum |r_n - s_n|$, or as the *sum* of *absolute errors*. The **L2 norm** $||E||_2$ is calculated as $\sqrt{\sum |r_n - s_n|^2}$, or as the *square-root* of the *sum* of *squared errors* (*euclidean distance* between the two vectors). The **L∞ norm** $||E||_\infty$ is calculated as $max(|r_n - s_n|)$, or as the *maximum* of *absolute errors*.

This experiment was for comparing the performance between PageRank computation with *L1, L2* and *L∞ norms* as convergence check, for *damping factor α = 0.85*, and *tolerance τ = 10^{-6}*. The *input graphs*, *system used*, and the rest of the *experimental process* is similar to that of the *first experiment*. Additionally, the execution time of each test case is measured using *std::chrono::high_performance_timer*. This is done *5 times* for each test case, and timings are *averaged (AM)*.

From the results, as seen in figures 3.1 and 3.2, it is clear that PageRank computation with **L∞ norm as convergence check is the fastest**, quickly followed by *L2 norm*, and finally *L1 norm*. Thus, when comparing two or more approaches for an iterative algorithm, it is important to ensure that all of them use the same error function as convergence check (and the same parameter values). This would help ensure a level ground for a good relative performance comparison.

Also note in figures 3.1 and 3.2 that PageRank computation with **L∞ norm** as convergence check **completes in a single iteration for all the *road networks* (ending with _osm)**. This is likely because it is calculated as *||E||_∞ = max(|r_n -*

$s_n|)$, and depending upon the *order (number of vertices) N* of the graph (those graphs are quite large), the maximum rank change for any single vertex does not exceed the *tolerance τ* value of $10^{-6}$.

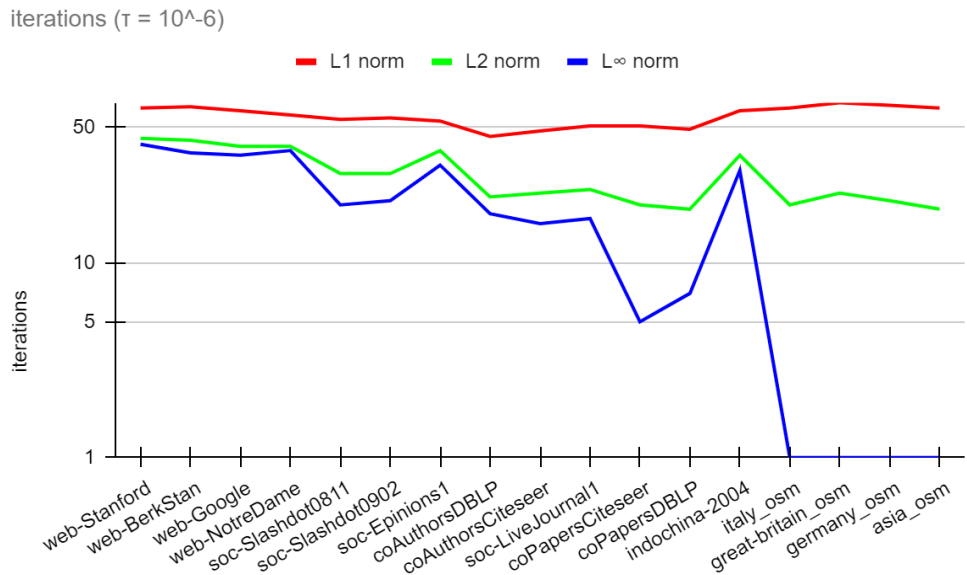

Figure 3.1: Iterations needed for PageRank computation for various graphs with *damping factor α = 0.85*, and *tolerance τ = $10^{-6}$*, with *L1*, *L2*, and *L∞ norms* as *convergence check*.

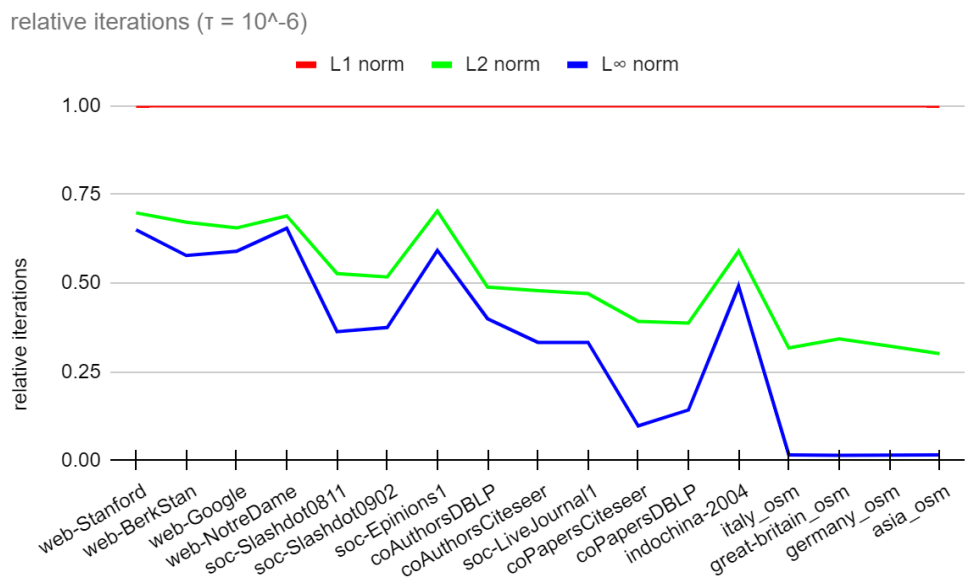

Figure 3.2: Relative iterations for PageRank computations for various graphs with *L1*, *L2*, and *L∞ norms* as *convergence check*, with respect to *L1 norm* as *baseline* (*damping factor α = 0.85*, and *tolerance τ = $10^{-6}$*).

| | | | | | | | | | |
|---|---|---|---|---|---|---|---|---|---|
| AM-RATIO | 4,645 ms | 2,052 ms | 1,070 ms | 1.00 | 0.44 | 0.23 | 4.34 | 1.92 | 1.00 |
| GM-RATIO | 1,291 ms | 626 ms | 236 ms | 1.00 | 0.48 | 0.18 | 5.46 | 2.65 | 1.00 |
| HM-RATIO | 351 ms | 198 ms | 118 ms | 1.00 | 0.56 | 0.34 | 2.98 | 1.68 | 1.00 |
| RATIO-AM | | | | 1.00 | 0.50 | 0.33 | 15.92 | 5.66 | 1.00 |
| RATIO-GM | | | | 1.00 | 0.48 | 0.18 | 5.46 | 2.65 | 1.00 |
| RATIO-HM | | | | 1.00 | 0.47 | 0.06 | 3.03 | 1.75 | 1.00 |

Table 3.1: Mean times for PageRank computation with L1, L2, and L∞ norms as convergence check (first 3 columns), ratios relative to L1 norm (second 3 columns), and then to L∞ norm (last 3 columns). Blue rows are unaffected by baseline choice, purple cells are close to ratios for mean iterations (see table 3.2), and red cells are significantly different.

| | | | | | | | | | |
|---|---|---|---|---|---|---|---|---|---|
| AM-RATIO | 57.29 | 28.82 | 18.94 | 1.00 | 0.50 | 0.33 | 3.02 | 1.52 | 1.00 |
| GM-RATIO | 56.91 | 27.50 | 10.21 | 1.00 | 0.48 | 0.18 | 5.57 | 2.69 | 1.00 |
| HM-RATIO | 56.51 | 26.32 | 3.55 | 1.00 | 0.47 | 0.06 | 15.91 | 7.41 | 1.00 |
| RATIO-AM | | | | 1.00 | 0.50 | 0.33 | 17.61 | 6.08 | 1.00 |
| RATIO-GM | | | | 1.00 | 0.48 | 0.18 | 5.57 | 2.69 | 1.00 |
| RATIO-HM | | | | 1.00 | 0.46 | 0.06 | 3.00 | 1.74 | 1.00 |

Table 3.2: Mean iterations for PageRank computation with L1, L2, and L∞ norms as convergence check (first 3 columns), ratios relative to L1 norm (second 3 columns), and then to L∞ norm (last 3 columns). Blue rows are unaffected by baseline choice, purple cells are close to ratios for mean iterations (see table 3.1), and red cells are significantly different.

In order to obtain a *composite relative performance ratio*, the six different methods mentioned above are calculated, for both PageRank computation *time* and *iterations*. They are listed in table 3.1 and 3.2 respectively. The ratios are calculated with both *L1*, and *L∞ norms* as *baseline*. Methods which calculate ratios at the end are not affected by the choice of baseline, as expected. However, not only is RATIO-GM unaffected as well, it is also identical to GM-RATIO, due to its mathematical property. It is observed that RATIO-AM and RATIO-HM are affected by the choice of baseline, while RATIO-GM is unaffected. RATIO-HM, RATIO-GM, and RATIO-AM with L1 norm as baseline are similar for both time and iterations, but HM-RATIO is significantly different. This indicates **GM-RATIO** to be the **most stable** composite relative performance ratio, **followed by AM-RATIO**. In fact, weighted GM-RATIO is used by SPECviewperf® [2], as mentioned above. Semantically, GM-RATIO comparison gives equal importance to the relative performance of each test case (graph), while an AM-RATIO comparison gives equal importance to magnitude (time/iterations) of all test cases (or simply, it gives higher importance to test cases with larger graphs).

## 4. Adjusting Tolerance

Similar to the *damping factor α* and the *error function* used for convergence check, **adjusting the value of tolerance τ** can have a significant effect. This experiment was for comparing the performance between PageRank computation with *L1, L2* and *L∞ norms* as convergence check, for various *tolerance τ* values ranging from $10^{-0}$ to $10^{-10}$ ($10^{-0}$, $5×10^{-0}$, $10^{-1}$, $5×10^{-1}$, …). The *input graphs*, *system used*, and the rest of the *experimental process* is similar to that of the *first experiment*.

For various graphs, some of which are shown in figures 4.1 and 4.2, it is observed that PageRank computation with *L1, L2,* or *L∞ norm* as *convergence check* suffers from **sensitivity issues** beyond certain (*smaller*) tolerance τ values, causing the computation to halt at maximum iteration limit (*500*) without convergence. As *tolerance τ* is decreased from $10^{-0}$ to $10^{-10}$, *L1 norm* is the *first* to suffer from this issue, followed by *L2 and L∞ norms (except road networks)*. This *sensitivity issue* was recognized by the fact that a given approach *abruptly* takes *500 iterations* for the next lower *tolerance τ* value.

It is also observed, as shown in figure 4.2, that PageRank computation with *L∞ norm* as convergence check **completes in just one iteration** (even for *tolerance τ ≥ $10^{-6}$*) for large graphs *(road networks)*. This again, as mentioned above, is likely because the maximum rank change for any single vertex for *L∞ norm*, and the sum of squares of total rank change for all vertices, is quite low for such large graphs. Thus, it does not exceed the given *tolerance τ* value, causing a single iteration convergence.

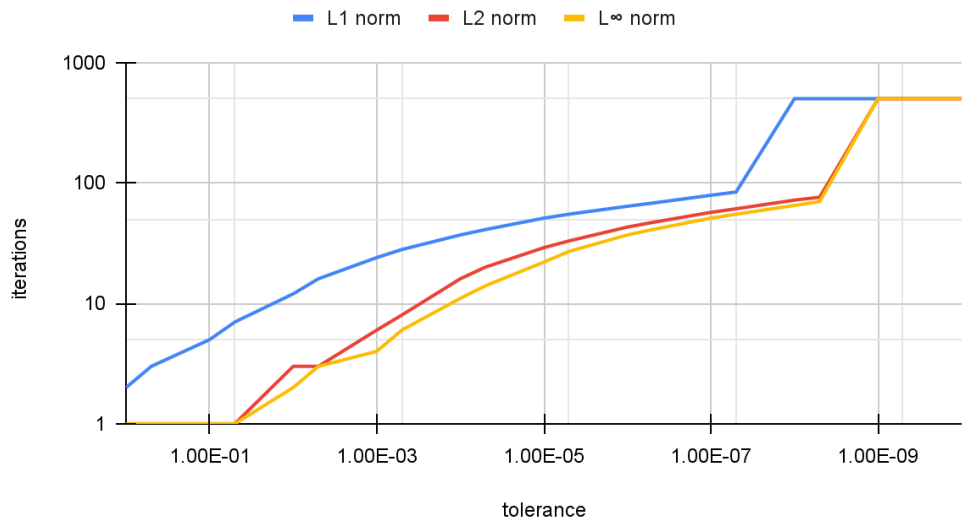

Figure 4.1: Iterations taken for PageRank computation of the *web-BerkStan* graph, with *L1*, *L2*, and *L∞ norms* used as convergence check. Up till *tolerance τ = 5×10$^{-2}$*, *L2 and L∞ norms* converge in just a single iteration. From *tolerance τ = 10$^{-8}$*, *L1 norm* begins to suffer from *sensitivity issues*, followed by *L2* and *L∞ norms* at *10$^{-9}$*.

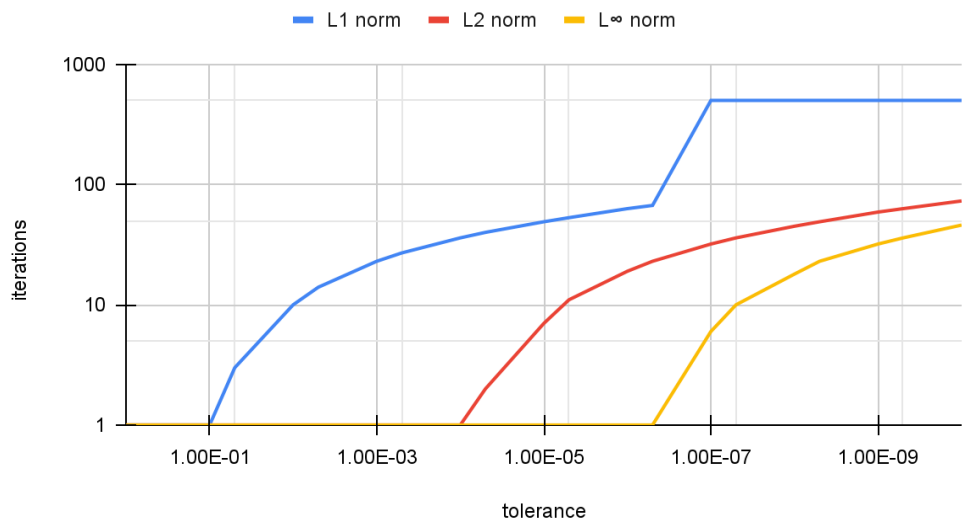

Figure 4.2: Iterations taken for PageRank computation of the *asia_osm* graph, with *L1*, *L2*, and *L∞ norms* used as convergence check. Until *tolerance τ = 10$^{-7}$*, *the L∞ norm* converges in just one iteration.

On average, PageRank computation with **L∞ *norm*** as the error function is the **fastest**, quickly **followed by *L2 norm***, and **then *L1 norm***. This is the case with both geometric mean (GM) and arithmetic mean (AM) comparisons of iterations needed for convergence with each of the three error functions, as shown in figures 4.3 and 4.4. In fact, this trend is observed with each of the individual graphs separately, although not shown here.

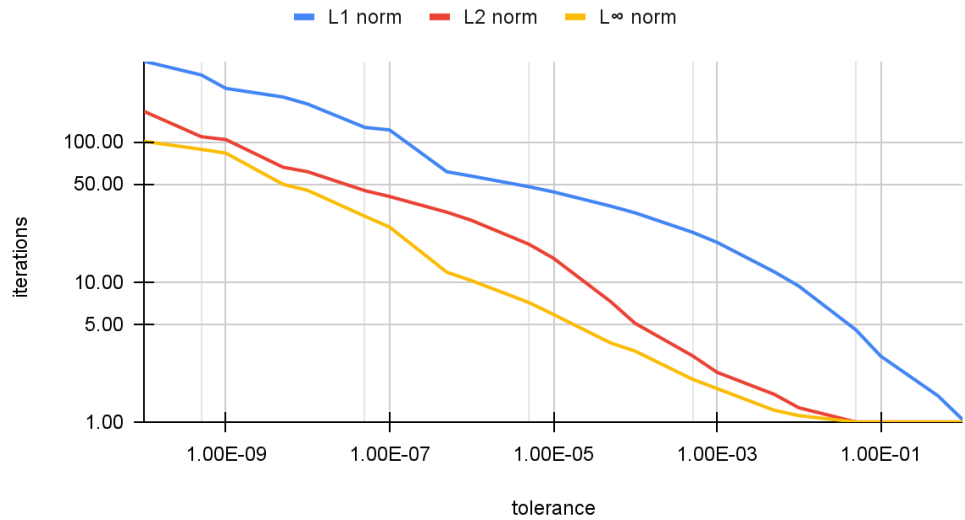

Figure 4.3: Geometric mean iterations taken for PageRank computation with *L1*, *L2* and *L∞ norms* as convergence check, and tolerance τ adjusted from *$10^{-0}$* to *$10^{-10}$*.

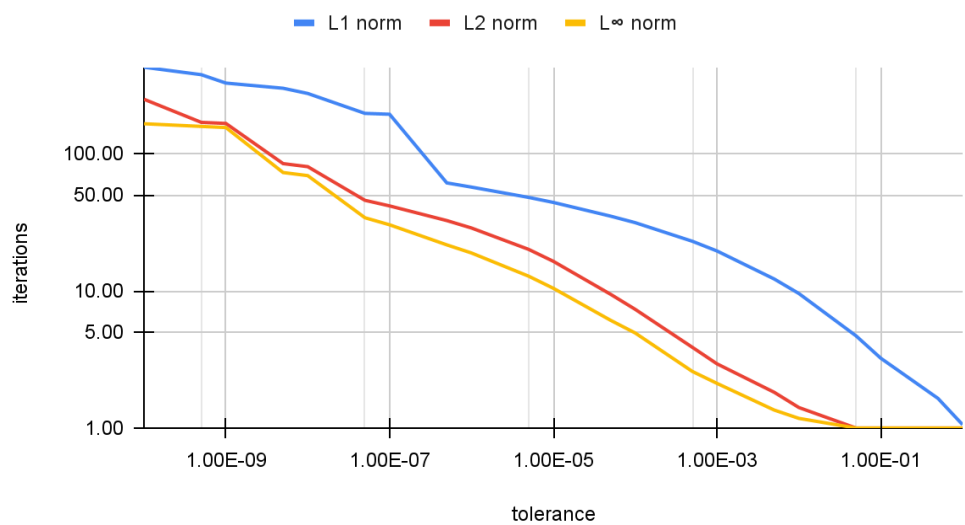

Figure 4.4: Arithmetic mean iterations taken for PageRank computation with *L1*, *L2* and *L∞ norms* as convergence check, and tolerance τ adjusted from *$10^{-0}$* to *$10^{-10}$*.

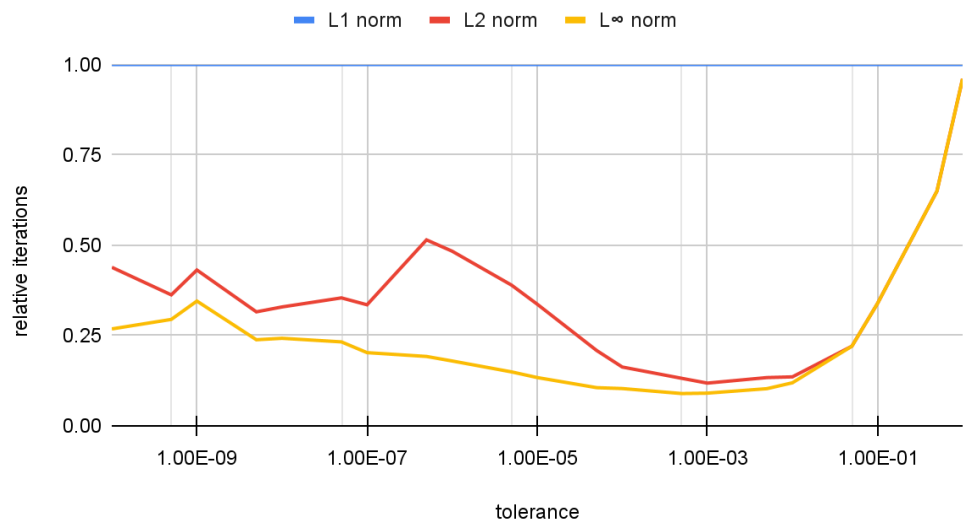

Figure 4.5: Relative GM iterations taken for PageRank computation with *L1*, *L2* and *L∞ norms* as convergence check, and tolerance τ adjusted from $10^{-0}$ to $10^{-10}$.

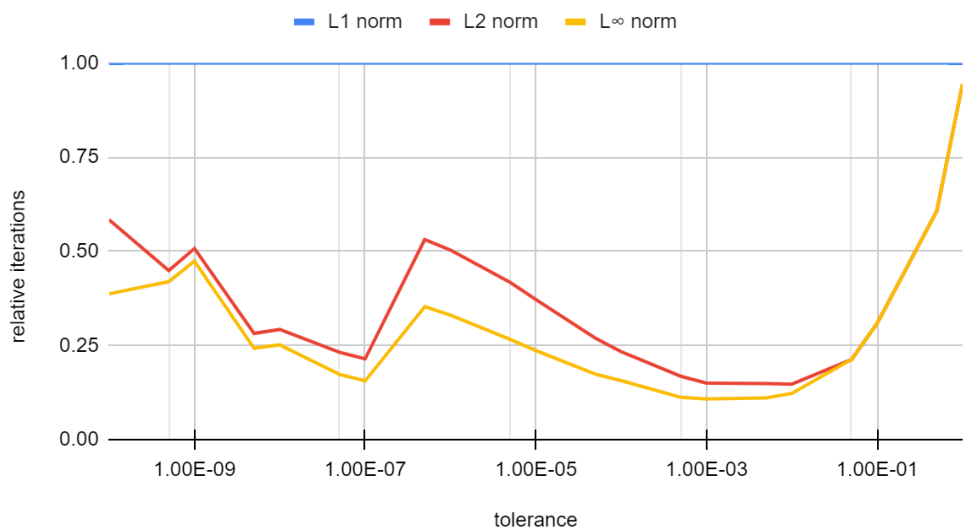

Figure 4.6: Relative GM iterations taken for PageRank computation with *L1*, *L2* and *L∞ norms* as convergence check, and tolerance τ adjusted from $10^{-0}$ to $10^{-10}$.

Based on **GM-RATIO** comparison, the *relative iterations* between PageRank computation with *L1*, *L2*, and *L∞ norm* as convergence check is **1.00 : 0.30 : 0.20**. Hence *L2 norm* is on *average 70% faster* than *L1 norm*, and *L∞ norm* is *33% faster* than *L2 norm*. This ratio is calculated by first finding the *GM* of *iterations* based on each *error function* for each *tolerance τ* value separately.

These *tolerance τ* specific means are then combined with *GM* to obtain a *single mean value* for each *error function (norm)*. The *GM-RATIO* is then the ratio of each *norm* with respect to the *L∞ norm*. The variation of *tolerance τ* specific means with *L∞ norm* as baseline for various *tolerance τ* values is shown in figure 4.5.

On the other hand, based on **AM-RATIO** comparison, the *relative iterations* between PageRank computation with *L1*, *L2*, and *L∞ norm* as convergence check is **1.00 : 0.39 : 0.31**. Hence, *L2 norm* is on *average 61% faster* than *L1 norm*, and *L∞ norm* is *26% faster* than *L2 norm*. This ratio is calculated in a manner similar to that of *GM-RATIO*, except that it uses *AM* instead of *GM*. The variation of *tolerance τ* specific means with *L∞ norm* as baseline for various *tolerance τ* values is shown in figure 4.6.

## 5. Conclusion

**Parameter values** can have a *significant effect* on performance, as seen in these experiments. Different **error functions** converge at *different rates*, and which of them converges faster depends upon the *tolerance τ* value. **Iteration count** needs to be checked in order to ensure that no approach is suffering from *sensitivity issues*, or is leading to a *single iteration convergence*. Finally, the **relative performance comparison** method affects which results get *more importance*, and which do not, in the *final average*. Taking note of each of these points, when comparing *iterative algorithms*, will thus ensure that the performance results are accurate and useful. The links to source code, along with data sheets and charts, for adjusting damping factor [5], error function [6], and tolerance [7] are included in references.